\documentclass[twocolumn,showpacs,preprintnumbers,amsmath,amssymb,prl]{revtex4}
\usepackage{graphicx}% Include figure files
\usepackage{dcolumn}% Align table columns on decimal point
\usepackage{bm}% bold math
\begin{document}
%************************************************************************
\title{Quantum effects on the phase diagram of nuclear-like systems}
\author{A. H. Raduta$^{1,2}$}
\affiliation{
  $^1$LNS-INFN Via S.Sofia 62, Catania \\
  $^2$NIPNE, RO-76900 Bucharest, Romania}
\begin{abstract}
A Path Integral Monte Carlo method is used to investigate the 
thermodynamics of nuclear like systems. Systems composed 
of bosons or fermions interracting via a Lennard-Jones potential with periodic 
boundary conditions were simulated and the corresponding phase diagrams are constructed. 
The Path Integral Monte Carlo appears to be a powerfull tool for investigating quantum effects 
in nuclear multifragmentation phenomena.
\end{abstract}

\maketitle

Highly excited nuclear systems are good ``laboratories'' for
thermodynamical studies. Due to the van der Waals type of the
nucleon-nucleon interaction these systems are supposed \cite{siemens}
to exhibit liquid-gas phase transitions.  The connection is however not 
straight forward due to their small number of constituents, the presence 
of the Coulomb force and quantum effects. The effect of the Coulomb 
interaction on the nuclear
liquid-gas phase transitions was previously studied (see e.g. Refs.
\cite{lee-mek,prl}). The effects of other multifragmentation specific features such as finite size (for attempts to describe the finite size effects of the system within Hartree-Fock theories see Refs. \cite{jaqaman,satpathy}) and degree of homogeneity were addressed in Refs. 
\cite{gross,bor,chomaz,phd}. 
Nuclear multifragmentation of excited nuclear systems
was treated with cluster-type models lyke SMM, MMMC, and MMM \cite{smm,mmmc,mmm}.
In this kind of models a fragmentation event is approached by various size
excited nuclear fragments interacting only via their Coulomb field. This approach
works very well at small densities where, the average distance between fragments
is large enough and fragments don't interact via nuclear forces. At larger densities,
fragments are nolonger spherica l and the nuclear interaction between them is quite 
important. At such densities, quantum and shape degeneracy effects play an important role.
Indeed, it was proven that a cluster-type statistical multifragmentation model cannot
accurately describe the dense branch of the phase diagram \cite{prl}.

Monte Carlo simulations with classical Lennard-Jones fluids showed that the accurate
treatment of fragments' shape degeneracy leads to the restoration of the Guggenheim shape of the 
phase diagram \cite{phd,ggnhm}. However, the systems considered there were classical. What
about quantum ones? This question is addressed in the present paper.

In the present work we perform a simulation of a quantum fluid interacting via a Lennard Jones
6-12 (LJ) fluid. Such simulations are tractable with Path Integral Monte Carlo (PIMC) methods
and have been extensively used in solid-state physics (see eg. \cite{Ceperley}).
The PIMC method is based on the isomorfism that one can acheive between the average value $\left<O\right>$ of 
an observable ($O$) of a N-particle quantum system and the average value of the the same observable correspnding 
to a classical N-polymer system. This correspondence was firstly pointed by Feynman in 1972 \cite{Feynman}.
Let us briefly present some Path Integral theory.
The average value of a system observable, $O$, writes:
\begin{equation}
\left<O\right>=\frac{1}{Z}\sum_i e^{-\beta E_i} \left< \Phi_i | O | \Phi_i \right>
\end{equation}
where $Z=\sum_i e^{-\beta E_i}$ is the canonical partition function $\Phi_i$ is the $i$th particle wave-function.
In a position-space representation, the density matrix writes:
\begin{equation}
\rho({\bf R},{\bf R'};\beta)=\left<{\bf R}| e^{-\beta E_i} |{\bf R'}\right>=\sum_i e^{-\beta E_i} \Phi_i^* \Phi_i
\end{equation}
where ${\bf R}=\{ {\bf r}_1, {\bf r}_2,...,{\bf r}_N \}$. So:
\begin{equation}
\left<O\right>=\frac{1}{Z}\int {\rm d}{\bf R} {\rm d}{\bf R'} \rho({\bf R},{\bf R'};\beta) \left< {\bf R} | O |{\bf  R'} \right>
\end{equation}
For a free particle in a $D$-dimensional box of size $L$ with periodic boundary conditions (PBC), one has:
\begin{equation}
\Phi_n=\frac{1}{L^{D/2}} e^{-i{\bf k}_n {\bf r}}.
\end{equation}
Hence:
\begin{equation}
\rho({\bf r},{\bf r'};\beta)=\frac{1}{L^D} \sum_n \exp\left[ -\beta \lambda {\bf k}_n^2 + i {\bf k}_n ({\bf r}-{\bf r}') \right] \nonumber
\end{equation}
\begin{equation}
= (4\pi \lambda \beta)^{-D/2} \exp\left[ \frac{({\bf r} - {\bf r}')^2}{4\lambda\beta}\right], {\rm if } \lambda \beta \ll L^2
\end{equation}
where for a particle of mass $m$, $\lambda=\hbar^2/2m$.
The underlying principle of introducing path integrals in {\em imaginary time} is the product property of the density matrix stating 
that the low temperature density matrix can be expressed as a product of high-temperature density matrices:
\begin{equation}
e^{-\beta {\cal H}} = \left(e^{-\tau {\cal H}}\right)^M
\end{equation}
where the ``time step'' $\tau=\beta/M$.
Usually the hamiltonian ${\cal H}$ is a sum of a kinetic part and a potential one: ${\cal H = K + V}$. For very small values
of $\tau$ one can use the so-called {\em primitive approximation} \cite{Raedt}:
\begin{equation}
e^{-\tau {\cal H}} = e^{-\tau {\cal K}} e^{-\tau {\cal V}}
\end{equation}
So, for very large values of $M$, one has:
\begin{equation}
e^{-\beta {\cal H}} \simeq \left( e^{-\tau {\cal K}} e^{-\tau {\cal V}} \right)^M
\end{equation}
The density matrix of a system of $N$ particles in the {\em primitive approximation} is given by the following path integral:
\begin{equation}
\rho({\bf R}_0,{\bf R}_M;\beta)=(4 \pi \lambda \tau)^{DNM/2} \int ... \int {\rm d} {\bf R}_1...{\rm d} {\bf R}_{M-1} \nonumber
\end{equation}
\begin{equation} 
\exp\left\{  \sum_{i=1}^M \left[ \frac{({\bf R}_{i-1}-{\bf R}_i)^2}{4\lambda\tau} + \frac{\tau}2 \left( V({\bf R}_{i-1})+V({\bf R}_i) \right) \right] \right\}
\end{equation}
The system's partition function writes:
\begin{equation}
Z=\iint {\rm d}{\bf R}_0 {\rm d}{\bf R}_M \rho({\bf R}_0,{\bf R}_M;\beta)
\end{equation}

Note that this partition function is similar to the partition function of a classical system of $N$ polymers. 
{\it Thus, evaluating the average value of any observable of the initial many-body quantum system is equivalent to evaluating
the average value of the observable in the classical system described by the partion function $Z$ given by eqs. (9), (10).}

So far we dealt with distinguishable particle wave functions. As we know the bosonic wave functions are symetric and the fermionic ones are antisymetric.
This writes:
\begin{equation}
\Phi_{B/F}({\bf R})=(\pm)^{\cal P} \Phi_{D} ({\cal P}{\bf R})
\end{equation}
where the indexes $B$ and $F$ stand for bosonic / fermionic and ${\cal P}$ is one of the $N!$ permutations between the particle labels of the 
$N$ particle the many-body coordinate ${\cal R}$. The sign $+$ corresponds to bosonic systems and the sign $-$ to the fermionic ones. One can obtain
the (anti)symetrization by applying the (anti)symetrization operators on the distinguishable particle wave-function:
\begin{equation}
\Phi_{B/F}({\bf R})=\frac1{\sqrt{N!}}\sum_{\cal P} (\pm)^{\cal P} \Phi_D({\cal P} {\bf R})
\end{equation}
so,
\begin{equation}
\rho_{B/F}({\bf R},{\bf R}';\beta)=\frac1{N!}\sum_{\cal P} (\pm)^{\cal P} \rho_D({\bf R}, {\cal P} {\bf R}';\beta)
\end{equation}
and, replacing $\rho_D$ with the corresponding path integral:
\begin{equation}
\rho_{B/F}({\bf R},{\bf R}';\beta)=\frac1{N!}\sum_{\cal P} (\pm)^{\cal P} \int ... \int {\rm d}{\bf R}_1...{\rm d}{\bf R}_{M-1} \nonumber
\end{equation}
\begin{equation}
\rho_D({\bf R},{\bf R}_1;\beta)...\rho_D({\bf R}_{M-1},{\cal P} {\bf R}';\beta)
\end{equation}

Now, having the bosonic and fermionic partition functions of the polymer-like system we can readily perform Metropolis Monte Carlo simulations in order to estimate 
average values of various observables. The principle is to generate a trajectory in the system's configuration space in agreement with the detailled ballanced 
principle. We will not insist here on the simulation since such methods are extensively explained elsewhere \cite{Ceperley}. However, it is worth noticing that 
when sampling fermions we deal with both positive and negative statistical wheights since the average value of any observable writes:
\begin{equation}
\left< O \right> = \frac{\sum_{\cal P} (-1)^{\cal P} \int {\rm d}{\bf R} {\rm d}{\bf R}'\left< {\bf R} | O | {\cal P} {\bf R}' \right> \rho ( {\bf R}, {\cal P} {\bf R}'; \beta)}
{\sum_{\cal P} (-1)^{\cal P} \int {\rm d}{\bf R} {\rm d}{\bf R}' \rho ( {\bf R}, {\cal P} {\bf R}'; \beta)}
\end{equation}
Therefore, we have the ratio between two differences. The problem arises at small temperatures where both differences are close to zero so that the statistical fluctuations 
increase dramatically. This effect is known as {\em the fermion sign problem}. However, since we perform simulations for highly excited nuclear-like systems the {\em fermion 
sign problem} shouldn't play here a big role.

Using the above exposed theoretical ingredients we perform Monte Carlo simulations for LJ bosonic and fermionic systems. 
The Lennard-Jones 6-12 (LJ) potential writes:
\begin{equation}
  v_0(r)=4\epsilon \left[\left(\frac{\sigma}{r}\right)^{12}-
    \left(\frac{\sigma}{r}\right)^6\right]
\end{equation}
We use the truncated and long-range corrected version of the above potential:
$v(r)=v_0(r)$ when $r<r_c$; $v(r)=0$ when $r \ge r_c$, corrections being subsequently
included in order to account for the effect of the neglected tail
\footnote{ Contributions from the neglected tail to the system's
potential energy and virial energy term
are taken respectively as:\\ $\Delta v=\frac12 \rho A
\int_{r_c}^{\infty} {\rm d}{\bf r}v(r)=8\pi \rho A
\epsilon[\sigma^{12}/(9r_c^9) -\sigma^6/(3r_c^3)]$; $\Delta
\mathcal{V}=\frac12 \rho A \int_{r_c}^{\infty} {\rm d}{\bf
r}(r~\partial v(r)/\partial r) =8\pi \rho A
\epsilon[4\sigma^{12}/(9r_c^9) -2\sigma^6/(3r_c^3)]$.}. 
We took $r_c=L/2$, $L$ being the size of the recipient, $\sigma=2.55$ fm
and $\epsilon=25.03$ MeV. These values are rather arbitrary but roughly describe the binding energies of various size nuclei. 
We simulated 16 quantum particles placed in a cubic recipient with periodic boundary conditions. 
Then, we construct the phase diagram of this system in both bosonic and fermionic quantics.
To this aim, we estimate pressure versus volume curves at constant temperature.
Pressure can be easily evaluated starting from its canonical definition: $P=T
\partial \ln Z(\beta,V)/\partial V$, where $Z(\beta,V)$ is the
system's canonical partition function as defined above. One gets the virial expression for pressure:
\begin{equation}
  P=\frac{N T}V-\frac1{3V}\left<\sum_{i<j} r_{ij} 
    \frac{\partial v(r_{ij})}{\partial r_{ij}}\right>,
\end{equation}
where $v(r_{ij})$ is the particle-particle interaction and $\left< \right>$ has the meaning of
canonical average.

%-------------------------figure 1 ------------------------------
\begin{figure}%[htb]
\vspace*{-.1cm} 
  \hspace*{-0.7cm} 
\includegraphics[height=9cm,angle=270]{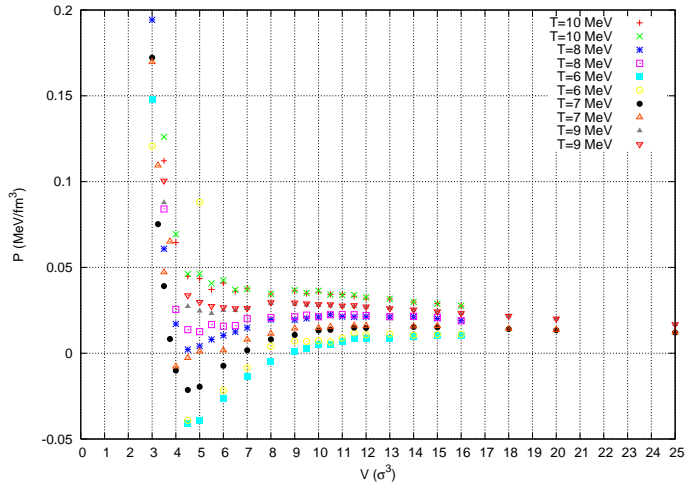}
\caption{Isothermal bosonic and fermionic presure versus volume curves corresponding to various temperatures for a 16 particle system.}
\end{figure}
%--------------------------------------------------------------------

Now let us discuss the results of the simulation. In Fig. 1 we represented pressure versus volume curves 
corresponding to both fermionic and bosonic 16 particle systems at temperatures ranging from 6 to 10 MeV.
One can observe the backbendings of these curves, signatures of a first order liquid-gas phase transition.
For a given temperature one can observe that all fermionic curves are above the bosonic ones. This can be 
better observed in Fig. 2 where bosonic and fermionic pressure versus volume curves corresponding to a temperature
of 7 MeV are represented.
One further performs Maxwell constructions on all collected pressure versus-volume curve and identify the liquid-gas
coexistence region for both fermionic and bosonic systems. This result is illustrated in Fig. 3. There one can 
observe the phase diagrams of both fermionic and bosonic systems. Some interesting aspects are to be noticed:
The fermionic coexistance region is wider than the bosonic one. The distance between the fermionic and the bosonic 
curves is larger on the denser branch of the phase diagram as expected since the quantm interaction between particles
is supposed to be larger in that region. Finally, the critical point of both bosonic and fermionic systems appears to be the same,
having the value of 10 MeV. This last point is particularly interesting showing that after the critical point the
quantum effects are no longer important, after that region both systems having a gas-like behaviour.
Taking into account the small size of the systems (A=16), the obtained criticall point temperature, is {\em quite good}. 
In infinite nuclear systems the critical temperature is supposed to be higher, critical temperature being smaller as the 
size of the system decreases \cite{phd}. Indeed, from Ref. \cite{phd} one can deduce a ratio of about 1.58 between the critical temperature
of an infinite system and the critical temperature of an A=16 system. One can therefore multiply the critical temperature
found herein of 10 MeV with the factor 1.58 and deduce the critical temperature of an infinite system, $T=15.8$ MeV which is 
a quite realistic value.

One can conclude that the Path Integral Monte Carlo is a powerfull tool for investigating highly excited nuclear systems. Though
the systems under consideration are rather small, still valuable information is obtained concerning the quantum effects on the 
systems' phase diagram. It was shown that both bosonic and fermionic have a common critical point. Moreover, the critical temperature value 
of 10 MeV obtainded for an A=16 system appears to be realistic, leading via extrapolation to a value of 15.8 MeV for infinite systems. 

%-------------------------figure 2 ------------------------------
\begin{figure}%[htb]
  \hspace*{-1.5cm} 
    \includegraphics[height=8cm,angle=270]{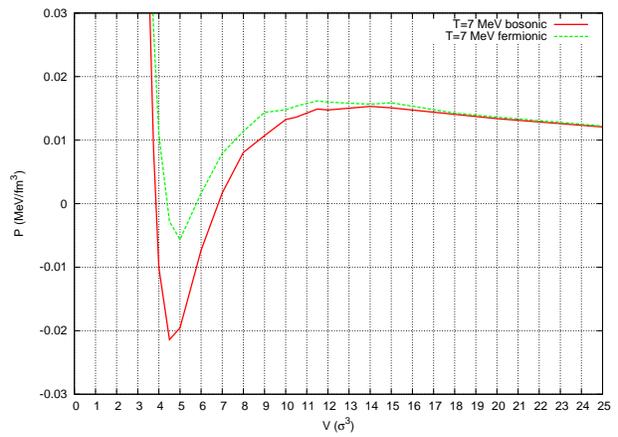}
  \caption{Isothermal bosonic and fermionic presure versus volume curves corresponding to a T=7 MeV temperature for a 16 particle system.}
\end{figure}
%--------------------------------------------------------------------
%-------------------------figure 3 ------------------------------
\begin{figure}%[htb]
  \hspace*{-1.5cm} 
    \includegraphics[height=8cm,angle=270]{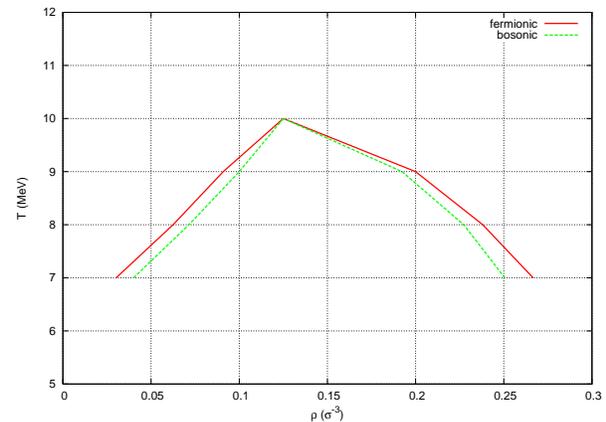}
  \caption{Phase diagram of a bosonic and a fermionic 16 particle in a box with PBC conditions system.}
\end{figure}
%--------------------------------------------------------------------

%%%%%%%%%%%%%%%%%%%%%%%%%%%%%%%%%%%%%%%%%%%%%%%%%%%%%%%%%%%%%%%%%%%%

\end{document}